\documentclass{article}

\usepackage{PRIMEarxiv}

\usepackage[utf8]{inputenc} % allow utf-8 input
\usepackage[T1]{fontenc}    % use 8-bit T1 fonts
\usepackage{hyperref}       % hyperlinks
\usepackage{url}            % simple URL typesetting
\usepackage{booktabs}       % professional-quality tables
\usepackage{amsfonts}       % blackboard math symbols
\usepackage{nicefrac}       % compact symbols for 1/2, etc.
\usepackage{microtype}      % microtypography
\usepackage{lipsum}
\usepackage{cite}
\usepackage{textcomp}
\usepackage{amsmath,amssymb,amsfonts}
\usepackage{tabularray} 
\usepackage{array}
\usepackage{color}
\usepackage{subcaption}
\usepackage{multirow}
\usepackage{fancyhdr}       % header
\usepackage{graphicx}       % graphics
\graphicspath{{media/}}     % organize your images and other figures under media/ folder
\usepackage{algorithm}
\usepackage{algpseudocode} 
\title{RICIAN DENOISING DIFFUSION PROBABILISTIC MODELS FOR SODIUM BREAST MRI ENHANCEMENT
}

\author{
  Shuaiyu Yuan\textsuperscript{1*}, Tristan Whitmarsh\textsuperscript{1} ,  Dimitri A Kessler\textsuperscript{1,2},  Otso Arponen\textsuperscript{1},  Mary A McLean\textsuperscript{1},\\ \textbf{Gabrielle Baxter}\textsuperscript{1}, \textbf{Frank Riemer}\textsuperscript{3}, \textbf{Aneurin J Kennerley}\textsuperscript{4,5}, \textbf{William J Brackenbury}\textsuperscript{4}, \\\textbf{Fiona J Gilbert}\textsuperscript{1} \textbf{and Joshua D Kaggie}\textsuperscript{1*} \\
  \textsuperscript{1}University of Cambridge, \textsuperscript{2}Universitat de Barcelona, \textsuperscript{3}Haukeland University Hospital,\\
  \textsuperscript{4}University of York,  \textsuperscript{5}Manchester Metropolitan University\\
  \textsuperscript{*}\texttt{\{sy442, jk636\}@cam.ac.uk} \\
}

\begin{document}
\maketitle

\begin{abstract}
Sodium MRI is an imaging technique used to visualize and quantify sodium concentrations in vivo, playing a role in many biological processes and potentially aiding in breast cancer characterization. Sodium MRI, however, suffers from inherently low signal-to-noise ratios (SNR) and spatial resolution, compared with conventional proton MRI. A deep learning method, the Denoising Diffusion Probabilistic Models (DDPM), has demonstrated success across a wide range of denoising tasks, yet struggles with sodium MRI's unique noise profile, as DDPM primarily targets Gaussian noise. DDPM can distort features when applied to sodium MRI. This paper advances the DDPM by introducing the Rician DDPM (RDDPM) for sodium MRI denoising. RDDPM converts Rician noise to Gaussian noise at each timestep during the denoising process. The model’s performance is evaluated using three no-reference image quality assessment metrics, where RDDPM consistently outperforms DDPM and other CNN-based denoising methods.  
\end{abstract}

% keywords can be removed

\section{Introduction}
Magnetic Resonance Imaging (MRI) is a widely used, non invasive, and radiation-free medical imaging technique for disease diagnosis. While MRI is mainly used for identifying disease locations, it can also be used for determining molecular and metabolic information without the need for an invasive biopsy through imaging techniques such as sodium MRI \cite{b1,b2}. Sodium MRI differs from conventional proton MRI, due to its inherent challenges of lower signal-to-noise ratios (SNR) and spatial resolution. These differences are caused by sodium’s lower concentration in the body and its lower gyromagnetic ratio compared to proton, affecting signal reception quality \cite{b3}. Deep learning (DL) denoising methods hold a significant potential to enhance sodium MRI image quality and offer new possibilities for clinical application, such as diseases characterization in breast cancer. 

In recent years, generative models, represented by Denoising Diffusion Probabilistic Models (DDPM) \cite{b4}, have been recognized as a powerful algorithm for enhancing medical image quality. The application of DDPM in medical image denoising has been studied through multiple modalities, demonstrating its versatility and effectiveness. For example, Gong et al. \cite{b5} employed the DDPM to denoise low-dose PET images, achieving better performance than other DL denoising methods. In the realm of MRI, Chung et al. \cite{b6} developed a regularized diffusion model to denoise and super-resolve knee and liver MR images, achieving state-of the-art results. Xiang et al. \cite{b7} proposed a self-supervised diffusion model to denoise diffusion-weighted MRIs and demonstrated superior denoising performance.  

Although the standard DDPM, as well as its variants, have achieved great success in denoising tasks across diverse imaging modalities, applying them to sodium MRI is challenging. A key characteristic of MRI is that its signals arise from complex data, which has Rician noise, especially at low SNR with magnitude-only data. In DDPM, however, there is a mismatch between the Gaussian noises assumed by the model, and the actual Rician noises present in MRI data. Directly applying DDPM to sodium MRI data may not adequately capture these non-Gaussian noise characteristics, potentially resulting in artifacts or distortions in the denoised images. Additionally, sodium MRI lacks high-quality reference images and large datasets, both of which are commonly expected for training diffusion models effectively.  

This work proposes a Rician DDPM (RDDPM) to address these limitations, which converts the Rician noise to Gaussian noise at each timestep of the DDPM denoising process. To train the RDDPM, we first train a convolutional neural network (CNN) model to denoise sodium MR images; the denoised images are then used to train the RDDPM.  

\section{Method}
\begin{figure}[!t]
\centerline{\includegraphics[width=\columnwidth]{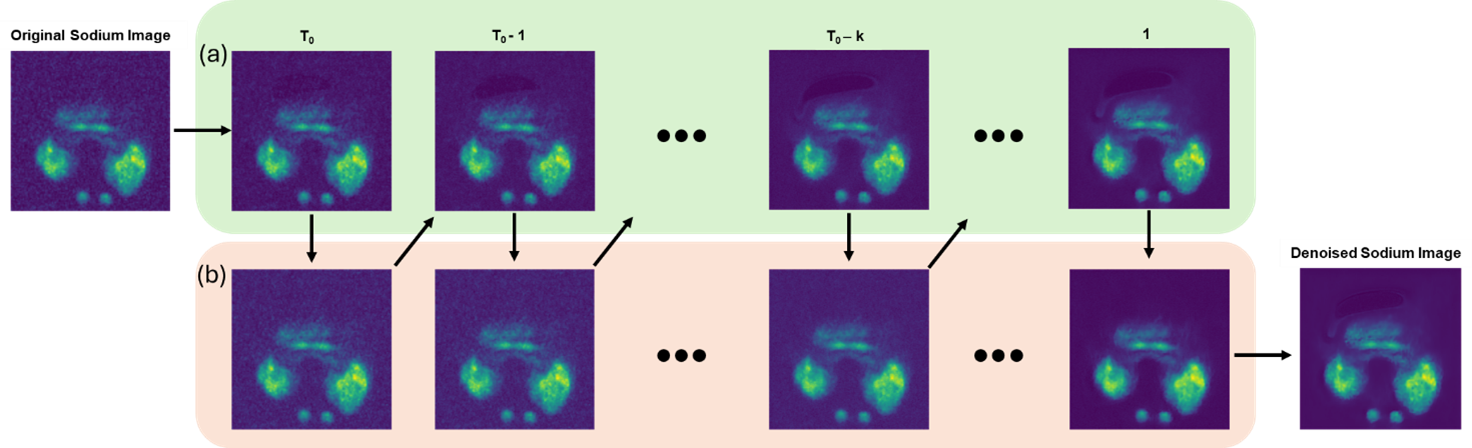}}
\caption{The pipeline of proposed RDDPM. (a) aims to convert the image with Rician noise to Gaussian noise at each timestep. (b) is a pre-trained DDPM.
}
\label{fig1}
\end{figure}
\subsection{Sodium MRI Denoising with CNN}
To denoise the sodium MR images with CNN models, such as Unet \cite{b8}, synthetic sodium MR images are created, which are formed by introducing sodium noise into T1-weighted (T1w) MR images \cite{b9} using the following equation,  
\begin{equation}R = \sqrt{\left(S + \frac{N}{\sqrt{2}}\right)^2 + \left(\frac{N}{\sqrt{2}}\right)^2}\nonumber\end{equation}
Where $R$ is the synthetic sodium MR image, $S$ is the T$_1$w MR image and $N$ is the sodium noise, which is the slices of the original sodium MRI data where there is no signal contained. The CNN models are then trained with pairs of T\textsubscript{1}w MR images and synthetic sodium MR images.   

\subsection{DDPM}
A DDPM is a generative model that learns to generate data by reversing a diffusion process. A DDPM is composed of two main stages: the forward (diffusion) process and the reverse (denoising) process \cite{b4}. The forward process involves gradually adding noise to a data sample $x_0$over $T$ timesteps to generate noisy samples $x_1$, $x_2$,…, $x_T$. This process can be defined as: 
\begin{equation}
q(x_t | x_{t-1}) = \mathcal{N}\left(x_t; \sqrt{1-\beta_t}x_{t-1}, \beta_t I\right)\nonumber
\end{equation}
where $\beta_t$ is a variance schedule that controls the amount of noise added at each timestep and $\mathcal{N}(.)$ denotes a Gaussian distribution. An alternative formulation directly computes a noisy sample $x_t$ at any timestep $t$ in a single step:
\begin{equation}
x_t = \sqrt{\bar{\alpha_t}} x_0 + \sqrt{1 - \bar{\alpha_t}} \epsilon
\nonumber
\end{equation}
where $\alpha_t = 1 - \beta_t$,  $\bar{\alpha_t}$ is the cumulative product of noise terms up to time $t$ and $\epsilon = \mathcal{N}(0,1)$.
The reverse process aims to recover the original data  by learning to gradually remove noise from  $x_t$ back to $x_0$. The reverse steps are parameterized by a neural network $\epsilon_\theta$, which predicts the distribution of  $x_{t-1}$ given $x_t$. This process is defined as:
\begin{equation}
x_{t-1} = \frac{1}{\sqrt{\bar{\alpha_t}}} \left( x_t - \frac{1 - \alpha_t}{\sqrt{1 - \bar{\alpha_t}}} \epsilon_\theta \right) + \sigma_t \epsilon
\nonumber
\end{equation}
where  $\sigma_t$ is the standard deviation of added noise. 

\subsection{Sodium MRI Denoising with RDDPM}
Unlike the Gaussian noise in the DDPM, which is zero mean and has the closure property, simply replacing the Gaussian noise in the DDPM with a Rician noise can collapse the model. In RDDPM, we use two parallel DDPM models. The pseudocode of the denoising procedure of RDDPM is shown in Algorithm 1 below. 
\begin{algorithm}
\caption{RDDPM denoising procedure}
\begin{algorithmic}[1]
\Require Magnitude of noisy image $A_t$, start timestep $T_0$, a pre-trained DDPM model $i$
\For{$t = T_0, \dots, 1$}
    \State $z_i \sim \mathcal{N}(0, I)$
    \State $z_j \sim \mathcal{N}(0, I)$
    \State $\hat{x_t^2} = \theta(A_t^2, t)$
    \State $\hat{\epsilon_t} = \text{model}_i(\hat{x_t}, t)$
    \State $x_{t-1} = \frac{1}{\sqrt{\bar{\alpha_t}}}({\hat{x_t} - \frac{1 - \alpha_t}{\sqrt{1 - \bar{\alpha_t}}} \hat{\epsilon_t}})$
    \If{$t \neq 1$}
        \State $x_t = x_{t-1} + \sigma_t z_i$
        \State $A_{t-1} = \sqrt{x_{t-1}^2 + (\sigma_t z_j)^2}$
    \EndIf
\EndFor
\State \Return $x_0$
\end{algorithmic}
\end{algorithm}

The input to the model is a noisy MRI magnitude image, $A_t$, where  $T_0$ is the starting timestep of the reverse process. For each timestep $t$, a model, $\theta$, predicts the squared Gaussian-noise image, $x_t^2$, from the squared Rician-noise image, $A_t^2$. Afterwards, a pre-trained DDPM model, $modeli$, is employed to perform the denoising process, whose weights are frozen. For timesteps $t \neq 1$, two sets of noise variables, $z_i$ and $z_j$, drawn from a standard normal distribution $N(0,1)$, are added to the denoised image maintaining stochasticity and modeling the Rician noise characteristics. The denoising process proceeds iteratively until the final timestep $t=0$, at which point the denoised image $x_0$ is returned as the output. 

As shown in Fig 1, the denoising process is visually represented through the progressive refinement of noisy MRI data over time. The leftmost image corresponds to the input noisy sodium MRI data, $A_{T_0}$, while the rightmost image represents the final denoised sodium MR image $x_0$. Each intermediate step shows the image after the application of noise transform and denoising process at the corresponding timestep, reflecting the gradual noise removal.

The training procedure of RDDPM is also different from that of DDPM. The pseudocode of the training procedure of RDDPM is shown in Algorithm 2 below.
\begin{algorithm}
\caption{RDDPM training procedure}
\begin{algorithmic}[1]
\Require Dataset $d$, maximum diffusion length $T_m$, inner epoch number $p_i$
\Repeat
    \State $x_0 \sim d(x_0)$
    \State $t \sim \mathcal{U}(\{1, \dots, T_m\})$
    \State $\epsilon_j \sim \mathcal{N}(0, I)$
    \State $x_t = \sqrt{\bar{\alpha_t}} x_0 + \sqrt{1 - \bar{\alpha_t}} \epsilon_j$
    \For{$p = 0, \dots, p_i$}
        \State $\epsilon_j \sim \mathcal{N}(0, I)$
        \State $A_t = \sqrt{x_t^2 + \left(\sqrt{1 - \bar{\alpha_t}} \epsilon_j \right)^2}$
    \EndFor
    \State Take gradient descent step on 
     \State $||x_t^2 - \theta(A_t^2)||^2$
\Until{converged}
\end{algorithmic}
\end{algorithm}

The first five steps are identical to those in DDPM, where Gaussian noise is progressively injected at each sampled timestep. Afterwards, for each inner iteration $p$, additional noise $\epsilon_j$ is added to create the noisy magnitude image $A_t$. This step ensures that the model is trained to handle the specific characteristics of Rician noise. In RDDPM, the network $\theta$ is trained to predict the squared Gaussian-noise image, $x_t^2$, from the squared Rician-noise image, $A_t^2$, using the mean square error (MSE) loss. 

\subsection{Evaluation}
In the case of the sodium MRI denoising, Full-Reference Image Quality Assessment (FR-IQA) metrics, such as Peak Signal-to-Noise Ratio (PSNR) and Structural Similarity (SSIM) \cite{b10}, are not feasible to evaluate the quality of denoised sodium MR images due to the absence of high-quality reference images. Therefore, three Non-Reference Image Quality Assessment (NR-IQA) metrics, Blind/Referenceless Image Spatial Quality Evaluator (BRISQUE) \cite{b11}, Multi-Scale Image Quality Transformer (MUSIQ) \cite{b12} and From Patches to Pictures (PaQ2PiQ) \cite{b13}, are used to evaluate the quality of denoised sodium images. BRISQUE is a commonly used NR-IQA metric while MUSIQ and PaQ2PiQ have demonstrated substantial agreement with expert evaluations \cite{b9}. For comparison, this study also applies one classic denoising algorithm, BM3D \cite{b14}, along with three other CNN-based models, DnCNN \cite{b15}, ResUnet \cite{b16} and ADNet \cite{b17}, to denoise sodium MR images.

\section{Experiments}
\subsection{Data}
This study used data from a cohort of 27 breast cancer patients \cite{b18}, which was acquired with informed consent and the approval of the local ethics board. Data of each patient comprised a high-quality T\textsubscript{1}w MR image and a sodium MR image. The image intensities are scaled to a range between 0 to 1. Three datasets were built: \textbf{Synthetic }\textbf{S}\textbf{odium }\textbf{D}\textbf{ata}\textbf{set} consisting of 270 slices of synthetic sodium MR slices, generated from T\textsubscript{1}w MR images of nine patients; \textbf{S}\textbf{odium }\textbf{T}\textbf{raining }\textbf{D}\textbf{ata}\textbf{set} consisting of 540 slices of authentic sodium MR slices from thirteen patients; \textbf{S}\textbf{odium }\textbf{T}\textbf{esting }\textbf{D}\textbf{ata}\textbf{set} consisting of 270 slices of authentic sodium MR images from the original nine patients. 

\subsection{Implementation}
All DL models were implemented using Pytorch, trained and tested on an Nvidia A100 GPU. All the CNN denoising models were trained using the Synthetic Sodium Dataset with batch size of 10, learning rate of 0.001, epoch number of 250, MSE loss and the Adam optimizer. Both DDPM and RDDPM were trained using the denoising results of Unet on the Sodium Training Dataset, with a learning rate of 0.0002 and the Adam optimizer. DDPM was trained for 1000 epochs. For RDDPM training, maximum diffusion length  $T_m$ was set as 40, inner iteration number  $p_i$ was set as 50. When testing DDPM and RDDPM on the Sodium Testing Dataset, for simplicity, the start timestep  $T_0$ was set as a fixed value of 15.

\section{Results and Discussion}
\begin{figure}[!ht]
    \centering
    \begin{minipage}[b]{0.45\textwidth}
        \centering
        \includegraphics[width=\textwidth]{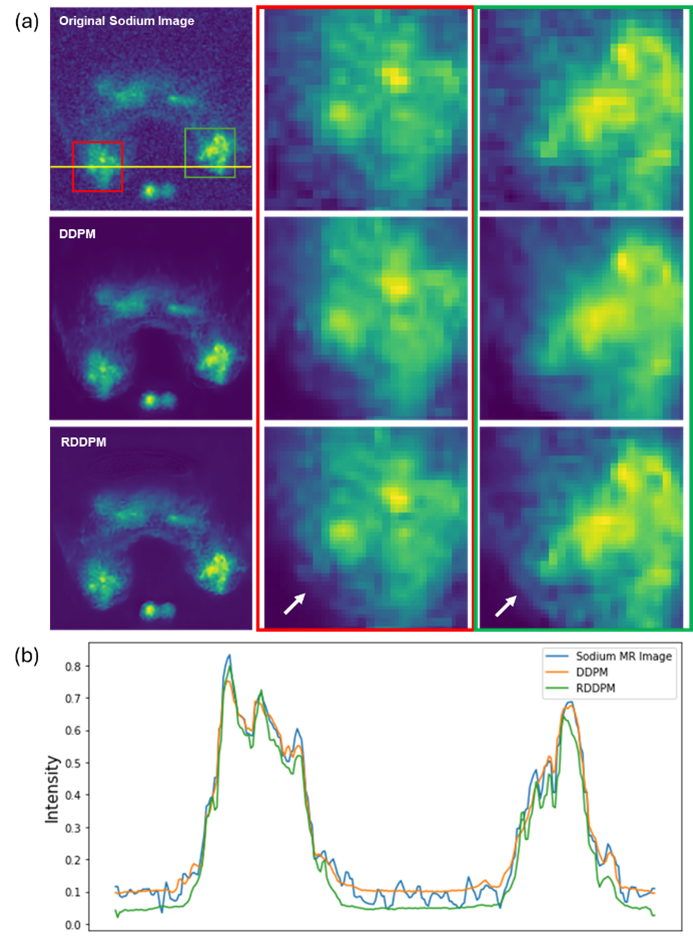} %
        \caption{An example of a sodium MR image prior to DL denoising, and the corresponding denoising results from DDPM and RDDPM. Two ROIs (a) and a yellow line (b) are selected and magnified for better comparison. }
        \label{fig2}
    \end{minipage}
    \hfill % This adds space between the subfigures
    \begin{minipage}[b]{0.45\textwidth}
        \centering
        \includegraphics[width=\textwidth]{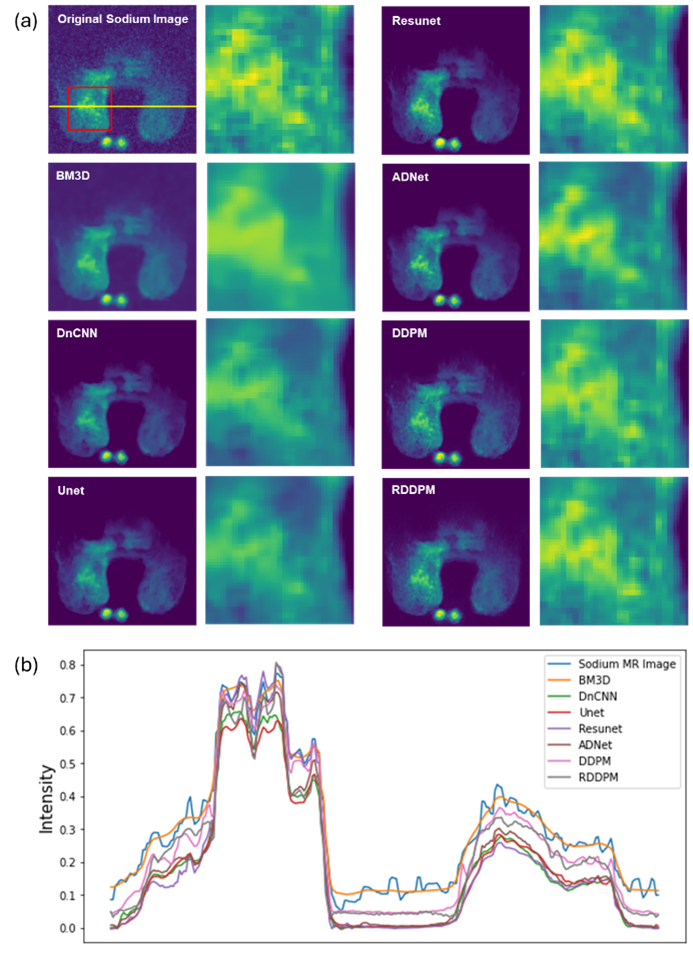} % Replace with your image file
        \caption{An example of a sodium MR image and the corresponding denoising results from BM3D, DnCNN, Unet, Resunet, ADNet, DDPM and RDDPM. A ROI (a) and a yellow line (b) are selected and magnified.}
        \label{fig3}
    \end{minipage}
    %\caption{Two figures side by side.}
    %\label{fig:sidebyside}
\end{figure}
\subsection{Denoising Results of DDPM and RDDPM}
Figure 2 presents an illustrative example comparing the original sodium MR image with the denoised outputs from DDPM and RDDPM. Two regions of interest (ROIs), marked by red and green squares, are magnified to improve visual clarity. A yellow line is also included to highlight the differences in the intensity profiles between the results of DDPM and RDDPM.

In Figure 2(a), compared to the original sodium MR image, DDPM introduces a higher degree of blurriness, while RDDPM preserves sharper edges and achieves better feature alignment with the original image. Figure 2(b) demonstrates that both DDPM and RDDPM perform well in aligning with the original image at high-SNR peaks. However, RDDPM shows superior performance in denoising low-SNR regions, effectively reducing background noise. 

\subsection{Comparison of Denoised Results}
\begin{table}[ht]
    \centering
    \scriptsize
    \caption{The image quality evaluation results of denoised sodium MR images. The best and the second-best results are emphasized in \textbf{Bold} and \textit{Italic}, respectively. }
        \centering
        \begin{tabular}{cccc }
        \hline 
        &BRISQUE& MUSIQ& PAQ2PIQ\\
        \hline 
        BM3D& 72.4820& 5.0743& 5.1764
\\ 
        DnCNN& 60.9133& 4.4017& 4.6366
\\ 
        Unet& 58.7705& 3.7178& 4.3825
\\ 
 Resunet& 52.7632& \textit{3.1747}&\textbf{4.2747}
\\
 ADNet& \textit{40.6889}& 3.6877&4.4727\\
 DDPM& 46.7097& 3.4669&4.3865
\\
 RDDPM& \textbf{34.4638}& \textbf{2.7866}&\textit{4.3806}
\\
\hline 
        \end{tabular}
\end{table}
Figure 3 presents the denoising results of BM3D, DnCNN, Unet, ResUnet, ADNet, DDPM and RDDPM. A ROI is highlighted by a red square, with a yellow line chosen for comparison. In Figure 3, BM3D demonstrated the poorest denoising performance, producing a blurred output and losing critical structural details. Both DnCNN and Unet blurred the image and substantially reduced the ROI's intensity. ADNet had sharper image quality than the DnCNN and Unet. From that CNN-based models generally underestimate the intensity of high-SNR peaks.

Table I lists the evaluation results of the denoised images on the Sodium Testing Dataset. The best and the second-best results are emphasized in \textbf{Bold} and \textit{Italic}, respectively. RPPDM achieves the best scores in BRISQUE and MUSIQ evaluations, and the second-best results in PAQ2PIQ, only second to Resunet. These results indicate the superior performance of RDDPM in denoising sodium MR images compared to other methods.

\section{Conclusion}
This paper introduces a novel model, Rician DDPM, designed to denoise sodium MR images. Through both qualitative visual assessments and three quantitative no-reference image metrics, RDDPM demonstrates superior performance compared to existing CNN-based methods. By addressing the inherent challenges of sodium MRI, RDDPM significantly improves image quality. When applied in more settings, this RDDPM denoised sodium MRI data may improve the molecular characterization of patient tumors, which could lead to an improvement in therapeutic strategies.

\section{Compliance with Ethical Standards}
The sodium MRI data used in this paper was obtained with ethical approval (IRAS ID: 260281; West Midlands - Black Country Research Ethics Committee) and informed consent. 

\section{Acknowledgements}
We acknowledge support from Cancer Research UK (A25922), the Medical Research Council (MR/X018067/1), CRUK Cambridge Centre and the NIHR Cambridge Biomedical Research Centre (BRC-121520014). This work was performed using resources provided by the Cambridge Service for Data Driven Discovery (CSD3) operated by the University of Cambridge Research Computing Service (www.csd3.cam.ac.uk), provided by Dell EMC and Intel using Tier-2 funding from the Engineering and Physical Sciences Research Council (capital grant EP/T022159/1), and DiRAC funding from the Science and Technology Facilities Council (www.dirac.ac.uk). 

%Bibliography

\end{document}